\documentclass{jfm}
\usepackage{graphicx}
\usepackage{epstopdf, epsfig}
\usepackage{natbib}
\usepackage{color}
\usepackage{amssymb}
\usepackage{amsmath}
\usepackage{bm} %bold math
\usepackage{latexsym}
\usepackage{lineno}
\newcommand\Reydels{\mathrm{Re}_{\delta_*}}

\newcommand\Reytau{\mathrm{Re}_{\tau}}
\newcommand\dd{\mathrm{d}}

\shorttitle{scaling of stream-wise Reynolds stress}

\shortauthor{P. A. Monkewitz}

\title{Reynolds number scaling and inner-outer overlap of stream-wise Reynolds stress in wall turbulence}

\author{Peter A. Monkewitz\aff{1}
  \corresp{\email{peter.monkewitz@epfl.ch}}}

\affiliation{\aff{1}\'Ecole Polytechnique F\'ed\'erale de Lausanne (EPFL), CH-1015, Lausanne, Switzerland}

\begin{document}

\maketitle   %Print title matter

%%%%%%%%%%%%%%%%%%%%%%%%%%%%%%%%%%%%%%%%%%%%%%%%%%%%%%%%%%%%%%%%%%%%%%
\begin{abstract}
The scaling of Reynolds stresses in turbulent wall-bounded flows is the subject of a long running debate.
In the near-wall ``inner'' region, a sizeable group, inspired by the ``attached eddy model'', has advocated the unlimited growth of $\langle uu\rangle^+$ and in particular of its inner peak at $y^+\approxeq 15$, with $\ln\Reytau$ \citep[see e.g.][and references therein]{smitsetal2021}. Only recently, \citet{chen_sreeni2021,chen_sreeni2022} have  argued on the basis of bounded dissipation, that $\langle uu\rangle^+$ remains finite in the inner near-wall region for $\Reytau\rightarrow\infty$, with finite Reynolds number corrections of order $\Reytau^{-1/4}$.
In this paper, the overlap between the two-term inner expansion $f_0(y^+) + f_1(y^+)/\Reytau^{1/4}$ of \citet{monkewitz22} and the leading order outer expansion for $\langle uu\rangle^+$ is shown to be of the form $C_0 + C_1\,(y^+/\Reytau)^{1/4}$. With a new indicator function, overlaps of this form are reliably identified in $\langle uu\rangle^+$ profiles for channels and pipes, while the situation in boundary layers requires further clarification. On the other hand, the standard logarithmic indicator function, evaluated for the same data, shows no sign of a logarithmic law to connect an inner expansion of $\langle uu\rangle^+$ growing as $\ln{\Reytau}$ to an outer expansion of order unity.\newline
\textbf{Submission history}: submitted April 22, 2023, to JFM Rapids; rejected August 26, 2023, after 18 weeks in the ``Rapids'' pipeline.
\end{abstract}

%The scaling of Reynolds stresses in turbulent wall-bounded flows is the subject of a long running debate.
%In the near-wall ``inner'' region, the large Reynolds number behavior of the peak stream-wise normal stress $\langle uu\rangle^+$ at $y^+\approxeq 15$ has divided the turbulence community. A large group, inspired by the ``attached eddy model'', advocates its unlimited growth with $\ln\Reytau$ \citep[see e.g.][and references therein]{smitsetal2021}, and a recent much smaller group has argued, on the basis of bounded dissipation, that near the wall $\langle uu\rangle^+$ remains finite for $\Reytau\rightarrow\infty$ and decreases from there as $\Reytau^{-1/4}$ \citep{chen_sreeni2021,chen_sreeni2022,monkewitz22}. Over the limited Reynolds number range, where good quality data are available, both asymptotic expansions provide reasonably close fits for the near-wall Reynolds stresses, in particular for the near-wall peak of $\langle uu\rangle^+$. This scaling issue is resolved in favor of $\langle uu\rangle^+$ remaining finite everywhere for all Reynolds numbers
%by analyzing the overlap, which links the inner region, where the variation of $\langle uu\rangle^+$ with $\Reytau$ is significant, to the outer region, where the variation is weak, of order $\mathcal{O}(\Reytau^{-1})$ or less.
%%%%%%%%%%%%%%%%%%%%%%%%%%%%%%%%%%%%%%%%%%%%%%%%%%%%%%%%%%%%%%%%%%%%%%
\section{\label{sec1}Introduction and outline of the problem}

Dividing wall-normal profiles of turbulence statistics in wall bounded flows into inner and outer parts connected through an overlap, is intrinsically a concept of matched asymptotic expansions \citep[abbreviated MAE, see e.g.][]{KC85,WilcoxP}. Its application to mean flow profiles can be traced back to the celebrated work of \citet{vonKarman34} and \citet{Millikan}, who introduced the logarithmic overlap law for the mean velocity profile.

Here and in the following, the classical non-dimensionalization is adopted with the ``inner'' or viscous length scale $\widehat{\ell} \equiv (\widehat{\nu}/\widehat{u}_\tau)$, and $\widehat{u}_\tau \equiv (\widehat{\tau}_w/\widehat{\rho})^{1/2}$, $\widehat{\rho}$ and $\widehat{\nu}$ the friction velocity, density and dynamic viscosity, respectively, with hats identifying dimensional quantities. The resulting non-dimensional inner and outer wall-normal coordinates are $y^+=\widehat{y}/\widehat{\ell}$ and $Y=y^+/\Reytau$, respectively, with $\Reytau\equiv \widehat{L}/\widehat{\ell}$ the friction Reynolds number and $\widehat{L}$ the outer length scale, i.e. the channel half height, pipe radius or boundary layer thickness.

Relative to the mean velocity, the situation for the Reynolds stresses is reversed, as the \textit{inner} parts vary significantly with $\Reytau$, while the outer parts quickly approach asymptotic profiles, with small finite Reynolds number corrections of order $\mathcal{O}(\Reytau^{-1})$ or less. In the present paper, the discussion focusses on the \textit{stream-wise Reynolds stress} $\langle uu\rangle^+$, as it is the component with the most data available.
%Other Reynolds stresses will be examined in a future, more exhaustive paper.

For this stream-wise component, the scaling of its inner part, and in particular of its inner peak height, is a subject of controversy. The two opposing views are summarized as follows:
\begin{itemize}
\item The Reynolds stresses scale according to the ``attached eddy'' model, in the following abbreviated ``AE'' model. Its main characteristic is the unbounded increase, proportional to $\ln{\Reytau}$, of Reynolds stresses in the inner near-wall region. This model has been
    first proposed by \citet{Towns56}, has recently been reviewed by \citet{MarusicMonty19} and has been extensively covered in the literature.
\item The Reynolds stresses remain finite in the limit of $\Reytau\!\to\!\infty$ everywhere in the flow. For the zero-pressure-gradient turbulent boundary layer, in the following abbreviated ZPG TBL, this view has been advanced by \citet{MonkNagib2015}. More recently, it has been further developed by \citet{chen_sreeni2021,chen_sreeni2022}, who have argued, on the basis of the ``law of bounded dissipation'', that in the inner, near-wall region, the finite Reynolds number corrections are of order $\mathcal{O}(\Reytau^{-1/4})$. This new scaling, in the following abbreviated as ``BD'' scaling for ``bounded dissipation'', has been taken up by \citet{monkewitz22}, who developed a composite asymptotic expansion for $\langle uu\rangle^+$, that compares well with a number of DNS and experimental data, but used an ad-hoc fit for the overlap and outer parts.\newline
\end{itemize}

The above alternative scalings correspond to the inner and outer asymptotic sequences
\begin{eqnarray}
\mathrm{BD\, scaling :}\ \{\Phi^{\mathrm{(in)}}_{\mathrm{BD}}\} =& \{1,\Reytau^{-1/4}, \Reytau^{-1}, ...\}\,; \ \{\Phi^{\mathrm{(out)}}_{\mathrm{BD}}\} = \{1,\Reytau^{-1}, ...\} \label{BD} \\
\mathrm{AE\, scaling :}\ \{\Phi^{\mathrm{(in)}}_{\mathrm{AE}}\} =& \{\ln{\Reytau}\, \&\, 1, \Reytau^{-1}, ...\}\,; \ \{\Phi^{\mathrm{(out)}}_{\mathrm{AE}}\} = \{1, \Reytau^{-1}, ...\} \label{AE}
\end{eqnarray}
where for both BD and AE scaling, the dependence of outer Reynolds stresses on $\Reytau$ is weak and finite Reynolds number corrections are thought to be of order $\mathcal{O}(\Reytau^{-1})$ at most. It is also noted, that the terms of order $\mathcal{O}(\ln{\Reytau})$ and of $\mathcal{O}(1)$ in equation (\ref{AE}) must, for the matching to the outer expansion, be treated together as a ``block'' \citep{CL73}, in the same way as for the matching of inner and outer mean velocity across the log law (see for instance \citet{MN2023}).

Discriminating between the two inner scalings on the basis of the $\Reytau$-dependence of the inner peak height $\langle uu\rangle^+_{\mathrm{IP}}$ at $y^+_{\mathrm{IP}} \approxeq 15$, has so far been inconclusive because of the limited Reynolds number range of reliable data. Both fits, $[a_1 \ln{\Reytau} + a_2]$ and $[b_1 + b_2\,\Reytau^{-1/4}]$ are defendable, as seen in \citet[][fig. 1]{Metal2010} and \citet[][fig. 2]{monkewitz22}, for instance. Determining the scaling of coefficients in the Taylor expansion of different stresses about the wall, as in  \citet{smitsetal2021}, is equally inconclusive for the same reasons. These authors have also challenged the use by \citeauthor{chen_sreeni2021} of the Taylor expansion of  $\langle uu\rangle^+$ about the wall to infer the scaling of the inner peak. The argument is valid insofar as using the Taylor expansion of $\langle uu\rangle^+$ across the region with the sharpest variation of the different terms in the transport equation for $\langle uu\rangle^+$ to estimate the \textit{magnitude} of its inner peak is questionable. However, this Taylor series argument is not relevant for the \textit{scaling} of the inner $\langle uu\rangle^+$ or any other quantity, as the scaling \textit{can only change across an overlap}. Clearly, no such overlap exists between the wall and the inner peak of $\langle uu\rangle^+$ at $y^+_{\mathrm{IP}} \approxeq 15$ !

The above synopsis of this scaling problem suggests, that the overlap between the inner and outer asymptotic expansions of $\langle uu\rangle^+$ has received insufficient attention. The overlap, also called common part, is a key element of MAE, which provides the smooth transition between inner and outer expansions based on  different asymptotic sequences, in particular the sequences (\ref{BD}) and (\ref{AE}). This is the subject of the next section \ref{sec2}, where the overlap is analyzed by determining, from channel DNS and experiments, the indicator function for the new BD scaling and comparing it to the standard log-indicator function for AE scaling.

In section \ref{sec3}, the two competing indicator functions are evaluated from experimental and DNS data for pipe flow, with results closely matching those for the channel. Section \ref{sec4} is then devoted to the ZPG TBL and reveals that the indicator functions are significantly different from channel and pipe, indicating a much faster drop of $\langle uu\rangle^+$ towards zero in the outer part of the boundary layer, presumably because of entrained free stream fluid. Before this drop, most ZPG TBL data appear slightly better fitted by the BD overlap, but the question remains open.

The conclusions in section \ref{sec5} are unequivocal in support of BD scaling for channel and pipe flow, while better data will be required to definitively settle the issue for the ZPG TBL.

\section{\label{sec2}The inner-outer overlap of $\langle uu\rangle^+$ for channel flow}

The inner-outer overlap plays a key role in MAE, as it smoothly connects inner and outer expansions. The choice tool to detect overlaps are indicator functions, well established for logarithmic overlaps. \newline

\textit{Logarithmic or AE overlap}\newline
Assuming that the inner part of $\langle uu\rangle^+$ scales according to the AE model (equation \ref{AE}), the functional form of the overlap is just the reverse of the mean flow overlap, where the $\ln{\Reytau}$ term is part of the outer expansion. Hence, the AE overlap of $\langle uu\rangle^+$  \textit{must} contain a term proportional to $\ln{(\Reytau /y^+)}\equiv -\ln{Y}$ in order to avoid a $\ln{\Reytau}$ term in the outer expansion. Indeed, in \citet{MMHS13}, authored by the principal advocates of the attached eddy model, the overlap law for $\langle uu\rangle^+$ is given as $-C \ln Y + D$ and the name of ``Townsend-Perry'' constant with a value of 1.26 was proposed for $C$.
However, this author is not aware of any published strong evidence for such a logarithmic law, such as a region of constant log-indicator function
\begin{equation}
\Xi_{\mathrm{AE}} = Y\left(\dd \langle uu\rangle^+/\dd Y\right) \equiv y^+\left(\dd \langle uu\rangle^+/\dd y^+\right) \quad .
\label{Xilog}
\end{equation}
There may have been the expectation that a region of constant  $\Xi_{\mathrm{AE}}$ would eventually develop at higher Reynolds numbers, but this is no longer tenable in view of the successful identification of the BD overlap.\newline

\textit{BD overlap}\newline
A composite expansion of $\langle uu\rangle^+$, based on the asymptotic sequence (\ref{BD}) has been constructed in \citet{monkewitz22}, in the following referred to as ``M22'', but the overlap and outer profiles were a bit of a ``bricolage'', which may be loosely translated from French as ``slapped together''. The construction consisted essentially of connecting the inner expansion at a fixed $y^+_\times = 470$ (equ. 2.7 of M22) to a fit of $\langle uu\rangle^+$ on the centerline.
The resulting logarithmic slope of the overlap (equ. 2.9 and section 3 of M22) was Reynolds number dependent, unlike the universal slope proposed by \citet{MMHS13} and others, but the choice of a logarithm in M22 may have been another manifestation of the long shadow of von K\'arm\'an.

\begin{figure}
\center
\includegraphics[width=0.48\textwidth]{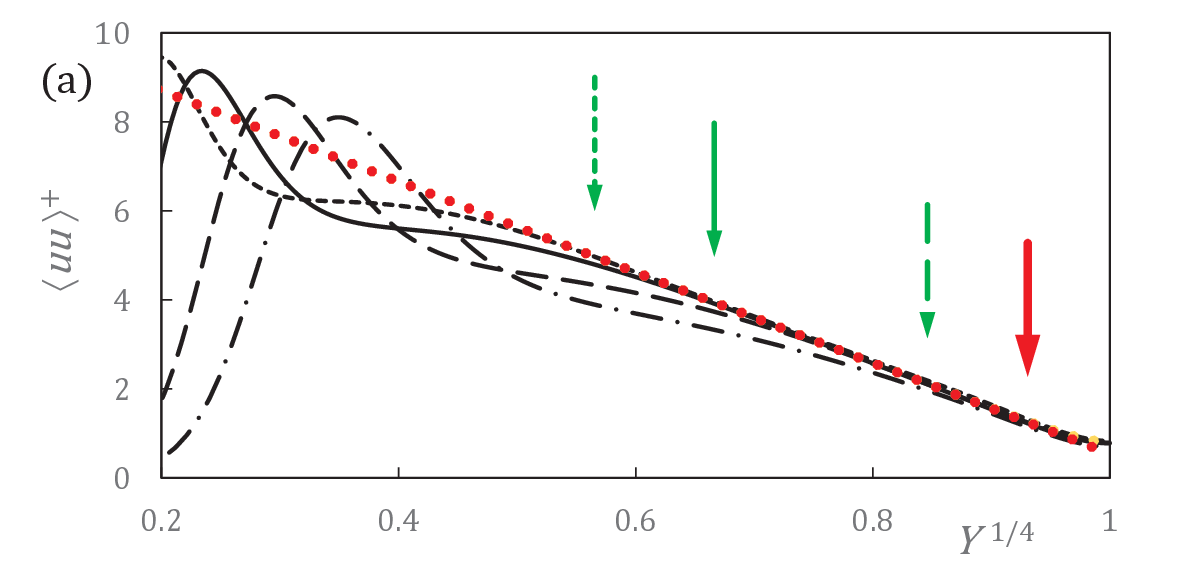}
\includegraphics[width=0.48\textwidth]{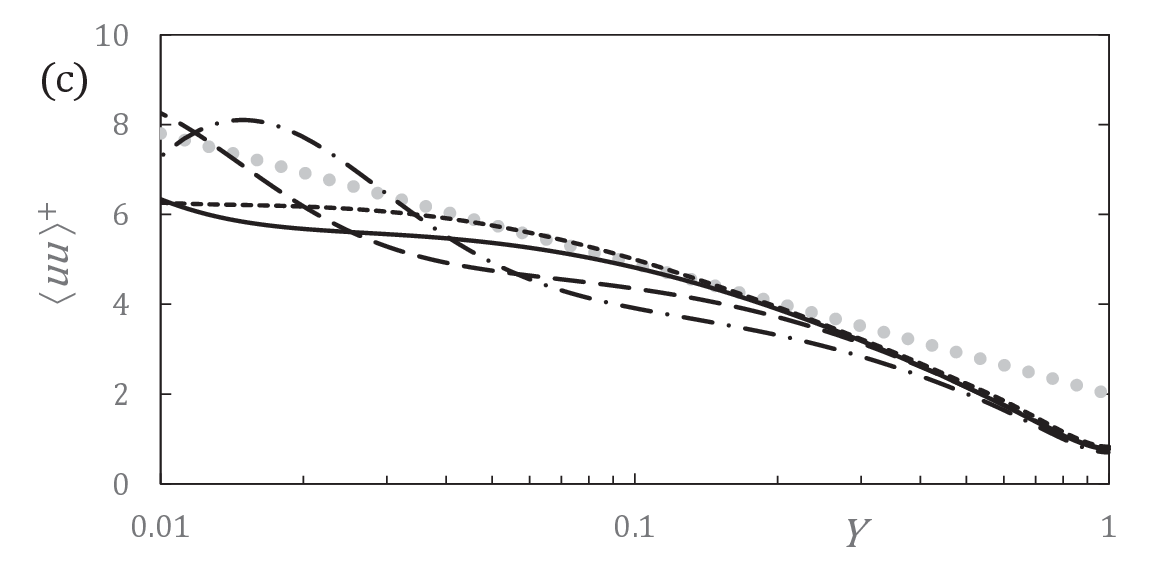}
\includegraphics[width=0.48\textwidth]{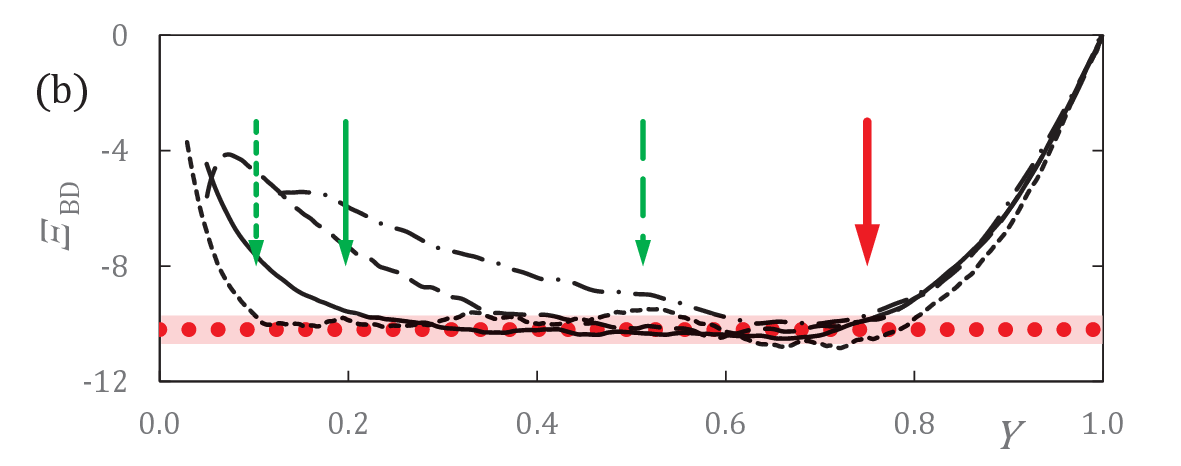}
\includegraphics[width=0.48\textwidth]{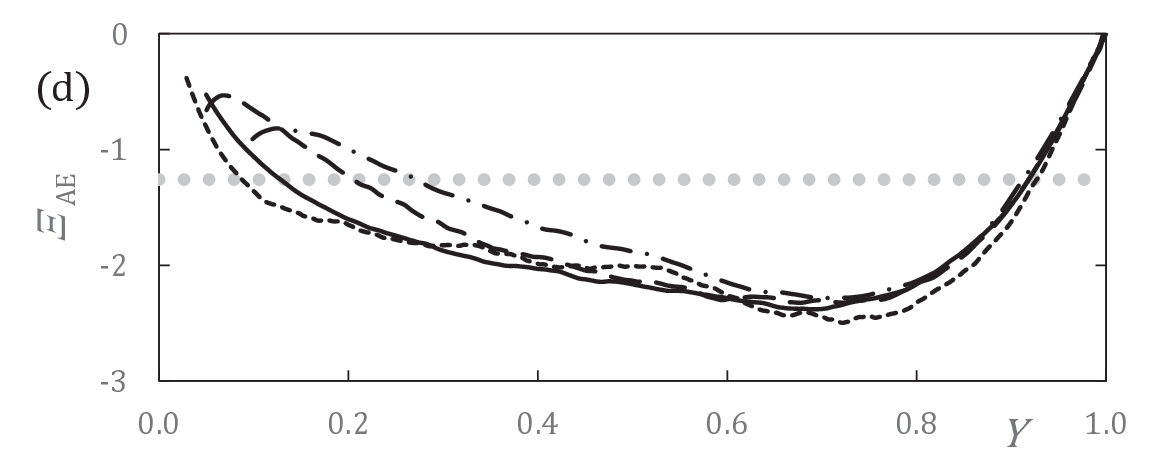}
\caption{\label{figuu} Inner-outer overlap laws of $\langle uu\rangle^+$ in channels: comparison of BD scaling (panels a \& b) and AE scaling (panels c \& d), with associated indicator functions for four channel DNS: $\Reytau = 10049$ \citep{HoyasOberlack2022} (short dashes), $\Reytau = 5186$ \citep{LM15} (solid line), $\Reytau = 1995$ \citep{LM15} (long dashes), $\Reytau = 1001$ (dash dots) \citep{LM15}.
(a) $\langle uu\rangle^+$ versus $Y^{1/4}$, with BD overlap of equation (\ref{uuCP}) (red $\bullet\bullet\bullet$); yellow $\bullet\bullet\bullet$, full outer fit (equation \ref{uuout}). Green vertical arrows: start of clean overlap at $y^+ \approx 10^3$ for $\Reytau$ = 10049, 5186 and 1995; red vertical arrow: end of clean overlap at $Y\approx 0.75$ ($Y^{1/4}\approx 0.93$)
(b) Corresponding BD indicator function (\ref{BDind}) versus $Y$. Red $\bullet\bullet\bullet$ : overlap value $-10.2$ of BD indicator, with red shaded area showing the range $-10.2 \pm 5\%$. Vertical arrows are at same $Y$-locations as in panel (a).
(c) Same $\langle uu\rangle^+$ as in panel (a) with AE overlap $(2 - 1.26\,\ln{Y})$ proposed by \citet{MMHS13} (grey $\bullet\bullet\bullet$). (d)
log-law indicator function (\ref{Xilog}) with proposed value of -1.26 (grey $\bullet\bullet\bullet$).}
\end{figure}

As it turns out, the overlap construction in M22 is just an awkward approximation of the new inner-outer overlap or common part (subscript ``cp'') of $\langle uu\rangle^+$ for BD scaling
\begin{equation}
\langle uu\rangle^+_{\mathrm{BDcp}}(Y) = 10.74 - 10.2\,Y^{1/4} \equiv 10.74 - 10.2\,(y^+)^{1/4}\,\Reytau^{-1/4} \label{uuCP}
\end{equation}
The complete leading order of the outer expansion is easily obtained by adding a ``wake'' to the overlap (\ref{uuCP})
\begin{equation}
\langle uu\rangle^+_{\mathrm{BDout}}(Y) = \langle uu\rangle^+_{\mathrm{BDcp}} + 0.26\,\exp\left(\frac{-10.2\,(1-Y)}{4 \times  0.26}\right) \label{uuout}
\end{equation}
Here, the coefficients in (\ref{uuCP})and (\ref{uuout}) have been determined from channel DNS, but will not be significantly different for the pipe.
Equations (\ref{uuCP}) and (\ref{uuout}) are seen in figure \ref{figuu}(a) to provide an excellent description of
the overlap of $\langle uu\rangle^+$ in channels and of the wake which becomes noticeable beyond $Y \approx 0.75$, marked by the red vertical arrow.

The indicator function to detect the new overlap law (\ref{uuCP}) is
\begin{equation}
\Xi_{\mathrm{BD}} = 4\, Y^{3/4}\,\big(\mathrm{d}\langle uu\rangle^+ \big/ \mathrm{d}Y \big) \equiv  4\,\Reytau^{1/4}\,(y^+)^{3/4}\,\big(\mathrm{d}\langle uu\rangle^+ \big/ \mathrm{d}y^+ \big)
\label{BDind}
\end{equation}
and is seen in figure \ref{figuu}b to develop a constant region for $\Reytau$'s beyond $10^3$, with a value of $\Xi_{\mathrm{BDcp}} = -10.2$. Furthermore, the region of constant $\Xi_{\mathrm{BD}}$ is seen to expand with increasing $\Reytau$, as expected from an overlap which ``starts'' at some $y^+_{\mathrm{startOL}}$ and ``ends'' at a  $Y_{\mathrm{endOL}}$, where it is understood, that the two ``boundaries'' depend on how much deviation of the full profile from the overlap is allowed.
The overlap start, defined here as $\Xi_{\mathrm{BD}}$ falling within $-10.2\,\pm\,5\%$, is shown for the three highest $\Reytau$ by vertical green arrows at $y^+_{\mathrm{startOL}} \approxeq 10^3$ in figures \ref{figuu}(a) and (b), while the end of the overlap, i.e. the location where the wake becomes significant, is located at $Y_{\mathrm{endOL}} \approxeq 0.75$, indicated by the vertical red arrow. It is noted in passing that the overlap (\ref{uuCP}) does of course not disappear around $\Reytau \approx 10^3$, but is ``buried'' under the inner and outer expansions, which move together as the inner-outer scale separation is reduced.

In figure \ref{figuu}(c) and \ref{figuu}(d), the same data are tested for the presence of a log-law, as proposed in \citet{MMHS13}, for instance. It is evident that the log-law indicator function (\ref{Xilog}) in figure \ref{figuu}(d) shows no sign of a plateau for the channel DNS analyzed. Hence the logarithmic fit of \citet{MMHS13} turns out to be an arbitrary tangent in figure \ref{figuu}(c).

\begin{figure}
\center
\includegraphics[width=0.6\textwidth]{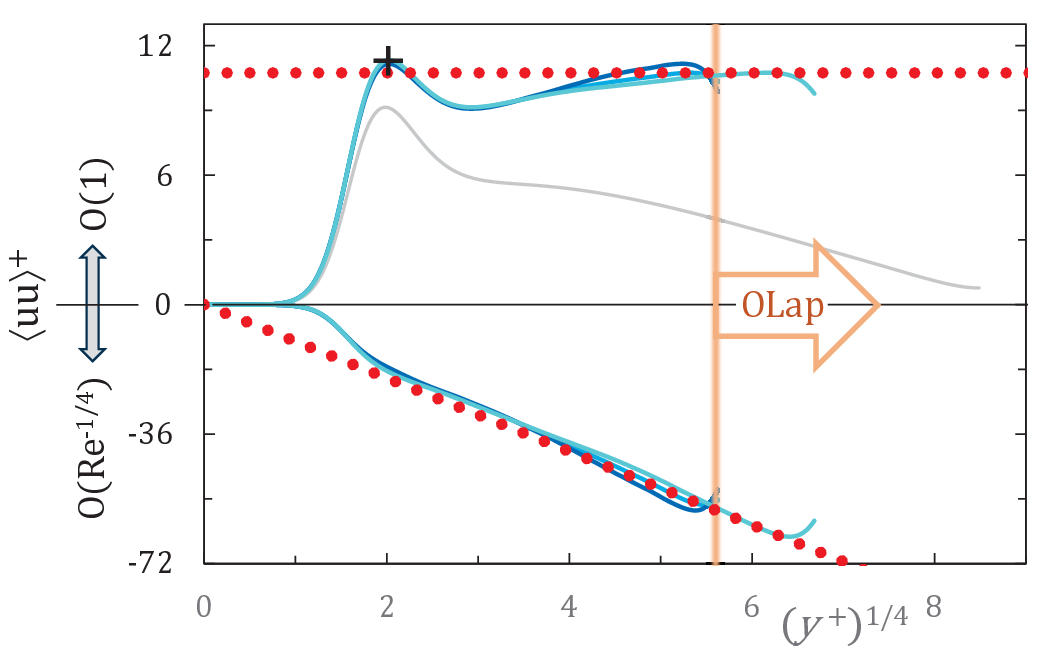}
\caption{\label{uuBDAE} The new BD overlap (\ref{uuCP}) of $\langle uu\rangle^+$ (red $\bullet\bullet\bullet$) versus $(y^+)^{1/4}$, with $\mathcal{O}(1)$ in the top part and $\mathcal{O}(\Reytau^{-1/4})$ in the bottom part of the figure. Approximate start of BD overlap, as defined in figure \ref{figuu}(a) and (b), indicated by vertical orange line at $(y^+)^{1/4}\approx 5.6$ ($y^+ \approx 10^3$). Blue lines: Two-term inner asymptotic BD expansion, determined in M22 from three pairs of channel DNS profiles (see also figure 1 of M22).
--- (grey), total $\langle uu\rangle^+$ from channel DNS of \citet{LM15} at $\Reytau = 5186 $ for comparison; + (black), limiting inner peak height $\langle uu\rangle^+_{\mathrm{IP}} = 11.3$ at $y^+ = 16.5$. }
\end{figure}

The above results are fully supported by figure \ref{uuBDAE}, which shows that the new overlap (\ref{uuCP}) is the appropriate large $y^+$ limit of the two-term inner expansion of $\langle uu\rangle^+(y^+) = f_0(y^+) + \Reytau^{-1/4}\,f_1(y^+)$, educed in M22 from three pairs of channel DNS profiles.

%In other words, (\ref{uuCP}) is the simplest large $y^+$ limit of the two-term inner expansion of $\langle uu\rangle^+ = f_0(y^+) + \Reytau^{-1/4}\,f_1(y^+)$, developed in M22, which is seen in fig. \ref{uuBDAE} to nicely asymptote to the new overlap (\ref{uuCP}), which smoothly connects to the leading order outer expansion.

\begin{figure}
\center
\includegraphics[width=0.48\textwidth]{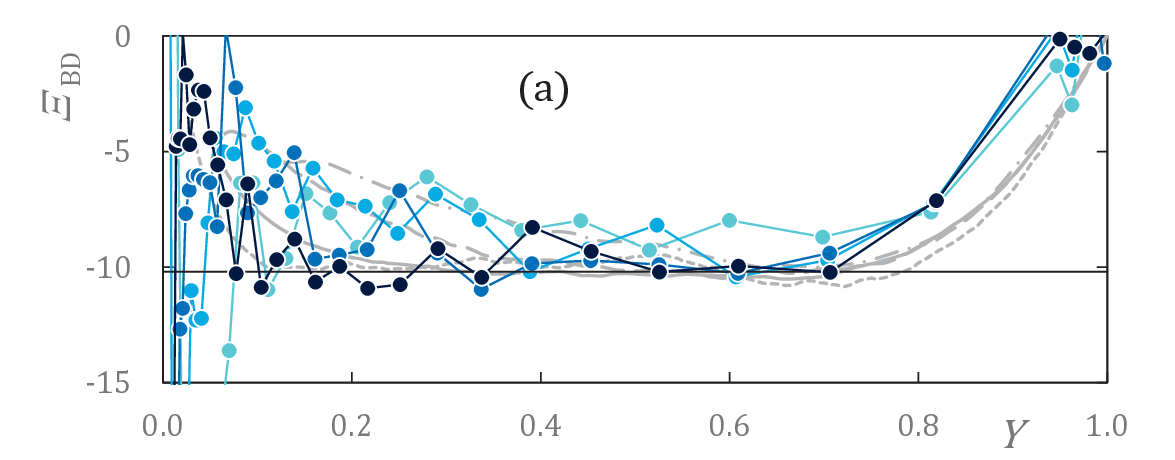}
\includegraphics[width=0.48\textwidth]{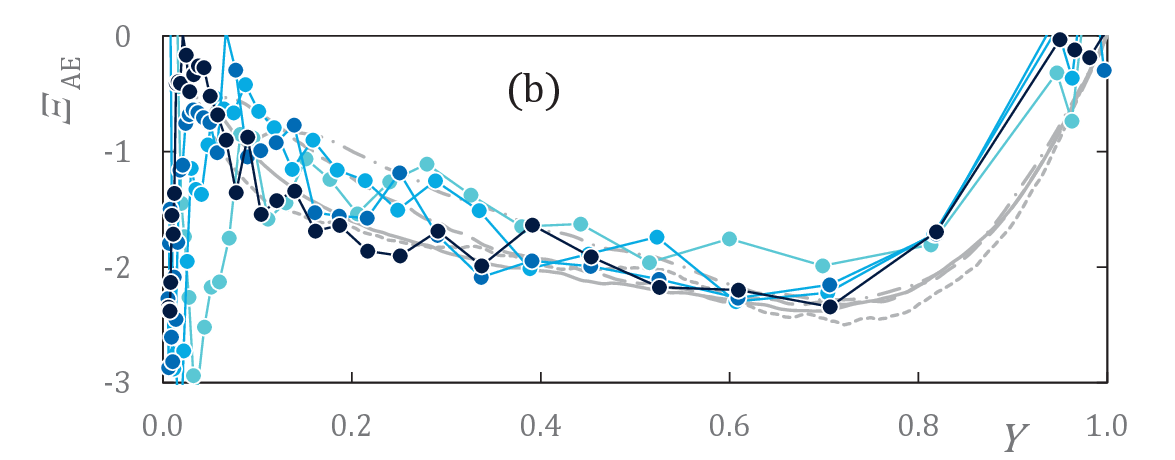}
\caption{\label{figXich} Indicator functions $\Xi_{\mathrm{BD}}$ (equation \ref{BDind}) in panel (a) and $\Xi_{\mathrm{AE}}$ (equation \ref{Xilog}) in panel (b) for the experimental channel data of \citet{SchultzFlack2013}, with $\Reytau = 1010, 1960, 4040, 5900$ (increasingly dark blue $-\bullet-$). Grey lines: reproduction, for comparison, of the four channel indicator functions in figures \ref{figuu}(b) and (d).}
\end{figure}

To reinforce the conclusion about the nature of the overlap deduced from channel DNS, the two indicator functions $\Xi_{\mathrm{BD}}$ and $\Xi_{\mathrm{log}}$ have been evaluated for the laser Doppler measurements of \citet{SchultzFlack2013} and are shown in figure \ref{figXich}. While there is considerable scatter due to the differentiation of experimental data, there can be no doubt that the data follow the bounded dissipation scaling, i.e. approach the same constant $\Xi_{\mathrm{BD}} = -10.2$ as the channel DNS in figure \ref{figXich}(b). The close correspondence between the experiment for $\Reytau=5900$ and the DNS for $\Reytau=5186$ is noted in particular.

%Further evidence for the BD-scaling in channels is the near perfect correspondence between the overlap (\ref{uuCP}) and the two-term inner expansion of $\langle uu\rangle^+$ educed from pairs of channel DNS in M22 and shown in figure \ref{uuBDAE}, which is figure 1 of M22 replotted against $(y^+)^{1/4}$. This figure establishes, that the inner $\langle uu\rangle^+(y^+)$ reaches the BD overlap (\ref{uuCP}) at $(y^+)^{1/4} \approxeq 5$, i.e. $y^+ \approxeq 600$.

\section{\label{sec3}The inner-outer overlap of $\langle uu\rangle^+$ for pipe flow}

\begin{figure}
\center
\includegraphics[width=0.48\textwidth]{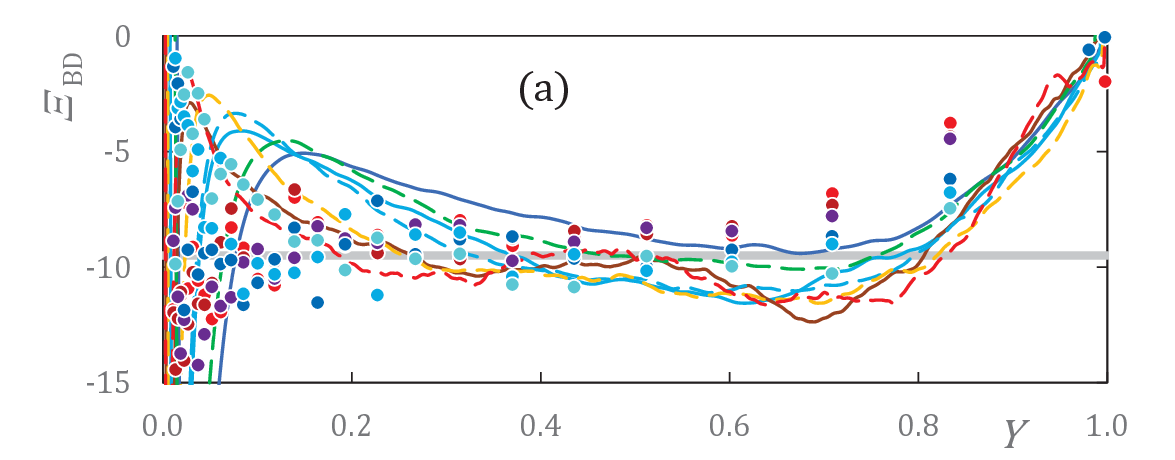}
\includegraphics[width=0.48\textwidth]{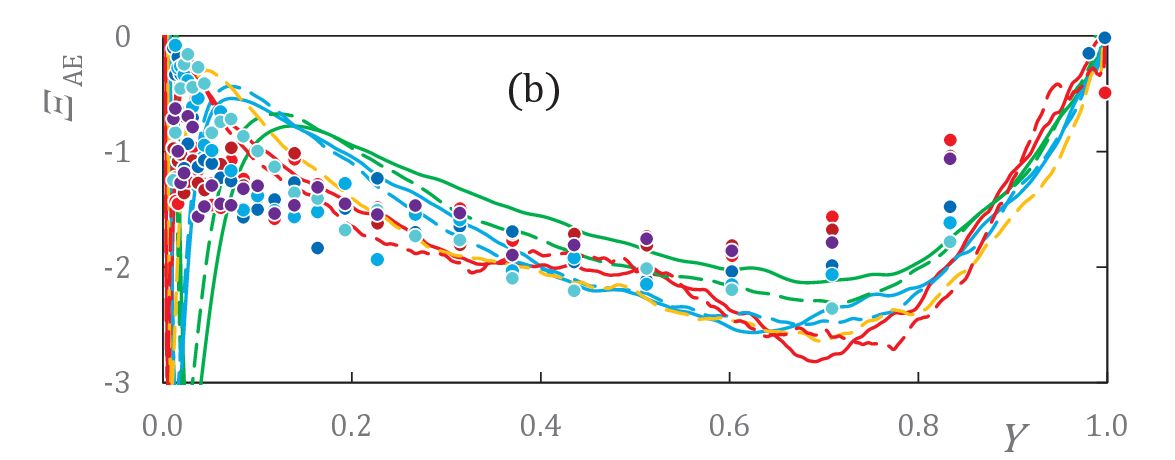}
\caption{\label{figpipe} Pipe flow: BD indicator function
(\ref{BDind}) in panel (a), compared to AE log-law indicator function (\ref{Xilog}) in panel (b). $\bullet$ (light, medium, dark blue, violet, dark red, red), smooth Superpipe data of \citet{Hultetal12} for $\Reytau = 5.41, 10.48, 20.25, 37.45, 68.37, 98.19 \times 10^3$; --- (green, blue, red), DNS data of \citet{Yao2023} for $\Reytau = 1.00, 2.00, 5.19 \times 10^3$; - - - (green, blue, orange, red), DNS data of \citet{pirozzoli_pipe2021} for $\Reytau = 1.14, 1.98, 3.03, 6.02 \times 10^3$. (grey) ---, estimated $\Xi_{\mathrm{BD}}=-9.5$ for the pipe overlap. }
\end{figure}

For pipe flow, $\Xi_{\mathrm{BD}}$ and $\Xi_{\mathrm{AE}}$ have been evaluated for the smooth Superpipe data of \citet{Hultetal12} and for selected DNS profiles of \citet{pirozzoli_pipe2021} and \citet{Yao2023}, all shown in figure \ref{figpipe}. As seen in panel (a), the data closely follow the BD overlap law of equation (\ref{BDind}) for the channel up to $Y \approx 0.4-0.5$ with $\Xi_{\mathrm{BDcp}}$ slightly increased from -10.2 to -9.5, marked by the grey horizontal line in figure \ref{figpipe}(a). Beyond $Y \approx 0.4-0.5$, the slope of $\langle uu\rangle^+$ in the pipe goes to zero faster than in the channel, due to the cylindrical geometry. Again, $\Xi_{\mathrm{AE}}$ in figure \ref{figpipe}b does not show any plateau, just as in figures \ref{figuu}(d) and \ref{figXich}(b) for the channel.

\section{\label{sec4}The overlap of $\langle uu\rangle^+$ in ZPG TBLs and its significant difference to channel and pipe}

\begin{figure}
\center
\includegraphics[width=0.48\textwidth]{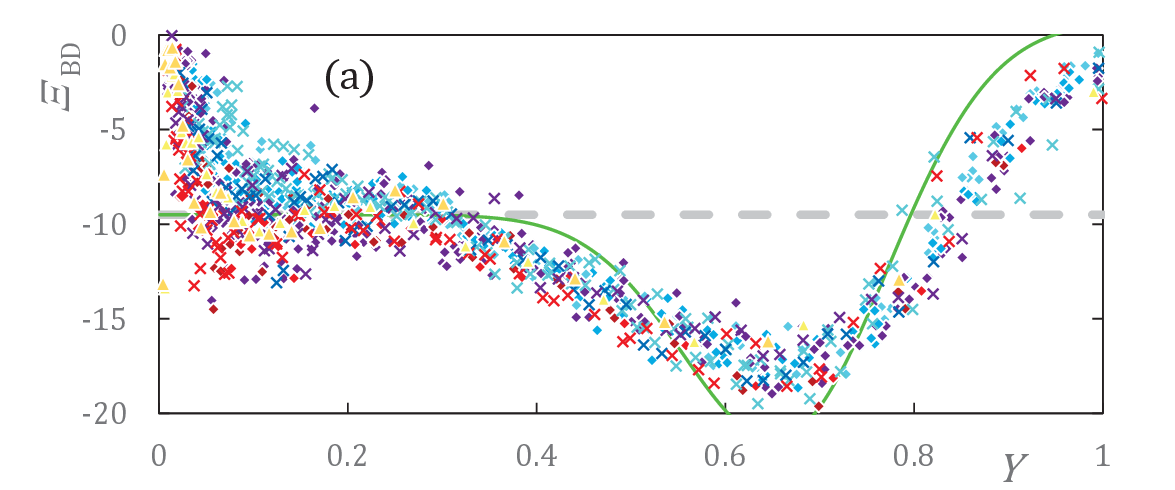}
\includegraphics[width=0.48\textwidth]{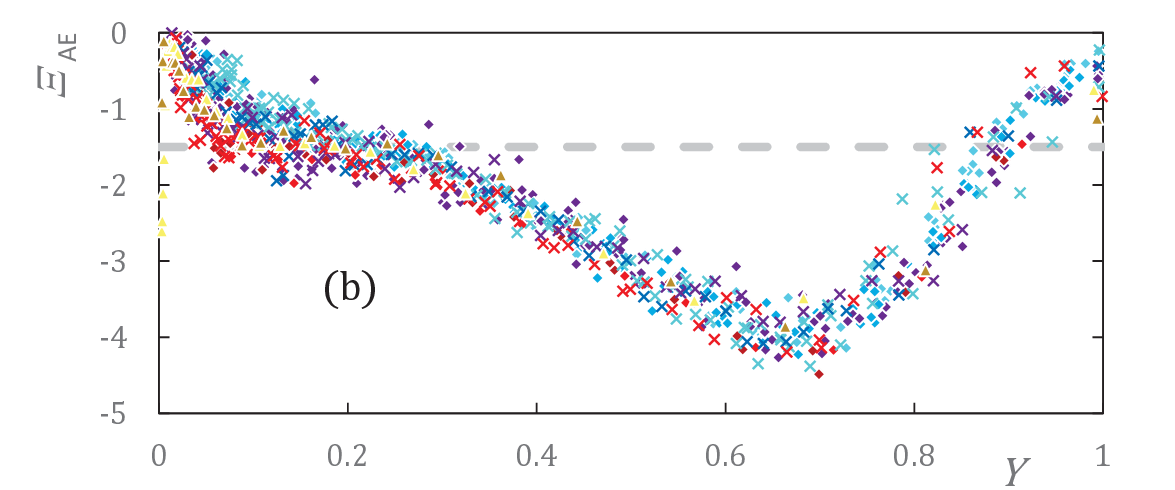}
\caption{\label{figZPG} ZPG TBL: BD indicator function
(\ref{BDind}) versus $Y$ in panel (a) compared to AE log-law indicator function (\ref{Xilog}) in panel (b). $\blacklozenge$, 22 profiles from \citet[][table 3, ref. ``IIT wind tunnel'' and ``I.Marusic, priv. comm.'']{MonkNagib2015} for $\Reydels \approx$ 15 and lower (light blue), 20 (blue), 30 (violet), 40 (dark red), 50 and higher (red) $\times 10^3$. $\times$, 7 profiles of \citet{KVickyphd} between $\Reydels \approx 10^4$ and $6.\,10^4$ (same color code). For these data, $Y = 3.5\,y^+/\Reydels$.
$\blacktriangle$ profiles of \citet{samie_etal_2018} for $\Reytau$ = $1.0\,10^4$ (light yellow), $1.45\,10^4$ (yellow), $2.0\,10^4$ (orange). (green) --- in panel (a), approximate intermittency correction of $\Xi_{\mathrm{BD}}$ according to equation (\ref{Xidecomp}).
(grey) - - -, possible level portions at $\Xi_{\mathrm{BD}} \approx -9.5$ in panel (a), and $\approx -1.5$ in panel (b).}
\end{figure}

The analogous comparison of BD and AE scaling for the ZPG TBL is shown in figure \ref{figZPG}. While the ZPG TBL is generally considered to be one of the ``canonical'' wall-bounded flows, both indicator functions are seen to be substantially different from the corresponding channel and pipe functions. Incidentally, the same is true for the ZPG TBL mean flow indicator function, which is also very different from channel and pipe, as shown in \citet{MN2023}.

When looking at the ``band'' of data in figure \ref{figZPG}, the
major differences to the channel and pipe indicator functions are evident:
\begin{itemize}
\item Both $\Xi$'s show a large negative excursion relative to channel and pipe in the range $0.2 \lessapprox Y \lessapprox 0.8$, indicating a much steeper decrease of $\langle uu\rangle^+$ in this region. A likely reason for this negative bulge is intermittency, as discussed below.
\item In the near-wall interval $0.1\lessapprox Y\lessapprox 0.25$
    one may see a short region of constant $\Xi_{\mathrm{BD}}\approx -9.5$ in figure \ref{figZPG}(a), while in panel (b) the data ``band'' in this region appears to have a slightly negative slope, but could equally well be fitted by a constant $\Xi_{\mathrm{AE}}\approx -1.5$. In short, the data scatter and the ``negative bulge'' beyond $Y\approx 0.25$ do not allow to discriminate between BD and AE scaling in the ZPG TBL, and the $\Reytau$'s of DNS are too low to help. However, it would be rather surprising, if the inner asymptotic sequence for $\langle uu\rangle^+$ in the ZPG TBL was different from channel and pipe!
\end{itemize}

To test the hypothesis that the large negative bulge of $\Xi$ beyond $Y\approxeq 0.25$ is due to intermittency, i.e. to the entrainment of free stream fluid, a rough model is developed, based on the location of the ``turbulent non-turbulent interface'' (TNTI). The PDF of its location has been studied by \citet[][fig. 3]{Chauhan14a}, who has determined its mean location $\bar{Y}_{\mathrm{TNTI}}=0.69$ and its standard deviation $\sigma=0.11$. With the cumulative distribution function $\mathcal{C}(Y)$ of the TNTI location , the measured $\langle uu\rangle^+$ may be expressed in terms of a hypothetical, fully turbulent stress as

\begin{equation}
\langle uu\rangle^+ = \langle uu\rangle^+_{\textrm{turb}}\,[1-\mathcal{C}(Y)] \,, \label{uudecomp}
\end{equation}
where the drastic simplification has been made, that $\langle uu\rangle^+ \equiv 0$ during the incursions of free-stream fluid into the boundary layer. With this, $\Xi_{\mathrm{BD}}$ can be decomposed as
\begin{equation}
\Xi_{\mathrm{BD}} = 4 Y^{3/4}\,\Big(\dd \langle uu\rangle^+_{\textrm{turb}} \Big/ \dd Y\Big) -  4 Y^{3/4}\,\Big(\dd [\mathcal{C}(Y)\,\langle uu\rangle^+_{\textrm{turb}}] \Big/ \dd Y\Big) \label{Xidecomp}
\end{equation}
Assuming that the first term in equation (\ref{Xidecomp}) corresponds to the hypothetical, non-intermittent ZPG TBL, with an overlap value close to the one for channel and pipe, the second term represents the intermittency correction. Concentrating on the overlap, the intermittency correction is evaluated with the channel overlap of equation (\ref{uuCP}) and the $\mathcal{C}(Y)$ of \citeauthor{Chauhan14a}. As shown in figure \ref{figZPG}(a), this model captures the essence of the deviation of $\Xi_{\mathrm{BD}}$ from the constant value $\approxeq -10$. It is furthermore noted, that the largest deviation of $\Xi_{\mathrm{BD}}$ from the channel and pipe indicator functions occurs at $Y\approxeq 0.65$, essentially at the mean location of the TNTI. This strongly supports the notion, that the difference between the ZPG TBL and the channel and pipe overlaps is principally due to the entrainment of free stream fluid.

\section{\label{sec5}Conclusions}

The clear conclusion from the present overlap analysis of the stream-wise Reynolds stress $\langle uu\rangle^+$ for channel and pipe flow is that $\langle uu\rangle^+$ remains finite everywhere in the limit of infinite Reynolds number and, in the inner region, decreases from there as $\Reytau^{-1/4}$. This has been demonstrated by analyzing the inner-outer overlap with the indicator functions $\Xi_{\mathrm{BD}}$ for the ``bounded dissipation'' scaling of \citet{chen_sreeni2021,chen_sreeni2022}, and comparing to $\Xi_{\mathrm{AE}}$ for the ``attached eddy'' or logarithmic scaling \citep[see for instance][]{MarusicMonty19}.
In other words, the unlimited growth of near-wall stream-wise Reynolds stress with $\ln{\Reytau}$ in channel and pipe flow is a feature of the attached eddy model and not of physical reality. One possible explanation for this result is the essentially inviscid nature of the attached eddy model. As $\ln\Reytau$ represents a weak divergence for $\Reytau\to\infty$, it may well be that it could be eliminated by introducing some viscous ``damping'' in the model, without compromising its physical attractiveness.

Indications have also been presented for BD scaling of $\langle uu\rangle^+$ in the ZPG TBL, but no definitive conclusion can be drawn on the basis of the available data. The main reasons are the short extent of the $\langle uu\rangle^+$ overlap, ending already at $Y\approxeq 0.25$, as opposed to $Y\approxeq 0.75$ in channels and pipes, combined with the relatively large scatter of the $\Xi$'s. As argued in section \ref{sec4},
the major difference beyond $Y\approxeq 0.25$ between the $\Xi$'s in the ZPG TBL and in channels and pipes is most likely due to the intermittent intrusion of free stream fluid, and sets the ZPG TBL clearly apart from channel and pipe.

In all likelihood, the present conclusions also apply to the other components of the Reynolds stress tensor, which will be the subject of a future full-length paper.
To conclude, a finite limit of Reynolds stresses for $\Reytau\!\to\!\infty$ is not only of theoretical interest, but has important technological implications, such as in ship hydrodynamics and hydraulic engineering.\newline

\begin{acknowledgments}
I am grateful to Katepalli Sreenivasan for his insightful comments and encouragement.\newline
\end{acknowledgments}

Declaration of Interests. The author reports no conflict of interest.
%%%%%%%%%%%%%%%%%%%%%%%%%%%%%%%%%%%%%%%%%%%%%%%%%%%%%%%%%%%%%%%%%%%%%%

\bibliographystyle{jfm}
\bibliography{Turbulence}

\end{document}